\documentclass[
 reprint,nofootinbib,
 amsmath,amssymb,
 aps,
]{revtex4-2}

\usepackage{graphicx}
\usepackage{dcolumn}
\usepackage{bm}
\usepackage{xcolor}
\usepackage[colorinlistoftodos]{todonotes}
\usepackage{soul}
\usepackage{setspace}
\usepackage{xcolor}
\usepackage[percent]{overpic}

\usepackage[colorlinks=true,
            linkcolor=blue,
            citecolor=blue,
            urlcolor=blue,
            ]{hyperref}

\begin{document}
\raggedbottom

\preprint{APS/123-QED}

\title{Data-driven modeling of shock physics by physics-informed MeshGraphNets}


\author{S.~Zhang$^1$}
\author{M.~Mallon$^2$}
\author{M.~Luo$^1$}
\author{J.~Thiyagalingam$^3$}
\author{P.~Tzeferacos$^4$}
\author{R.~Bingham$^{3}$}
\author{G.~Gregori$^1$}

\affiliation{%
 ${}^1$ Department of Physics, University of Oxford, Parks Road, Oxford OX1 3PU, UK
}%

\affiliation{%
 ${}^2$ Mechanical Department, Directorate of Technology, ESA-TEC, Keplerlaan 1, 2201 AZ Noordwijk, The Netherlands
}%


\affiliation{%
${}^3$ STFC Rutherford Appleton Laboratory, Didcot, Oxfordshire, OX11 0QX, UK. }

\affiliation{%
${}^4$ Department of Physics and Astronomy, University of Rochester, Rochester, New York 14627, USA. }


\begin{abstract}
High-resolution fluid simulations for plasma physics and astrophysics rely on Particle-in-cell (PIC) and hydrodynamic solvers (e.g., FLASH) to resolve shock-dominated, multi-scale phenomena, but their high computational cost severely limits scalability. This motivates the development of learning-based surrogate models, which offer a promising route to accelerate these simulations while preserving physical fidelity. In this work, we study the Sedov–Taylor shock propagation problem using a physics-informed graph-based surrogate model, Physics-Informed MeshGraphNet (Phy-MGN), designed for grid-based hydrodynamics. By incorporating weak physics constraints derived from the Euler equations using finite difference method, the model captures the self-similar shock evolution and associated flow structures without explicitly solving the full hydrodynamic equations at each timestep. Comparing to the baseline MeshGraphNet model, Phy-MGN is able to generalize beyond the training regime with a higher accuracy and preserves differentiability in parameter space while achieving a substantial reduction in computational cost relative to conventional numerical solvers.

\end{abstract}

\maketitle


\section{\label{sec:level1}Introduction}
Simulating high-resolution, multi-scale, and multi-physics plasma dynamics using traditional numerical methods is a complex and computationally demanding task. Particle-in-cell (PIC) codes~\cite{Dawson1983, birdsall2004plasma, Fonseca2002} and radiation-hydrodynamic codes (e.g., FLASH) \cite{Fryxell2000, Dubey2002} are widely used for modeling these phenomena and have been shown to accurately predict plasma behavior within their respective domain of validity. These high-fidelity simulators iteratively solve partial differential equations that describe the laws of physics~\cite{ brandstetter2023}, making the production of high-resolution multi-scale simulations extremely time-consuming. Simulations often require several days or even weeks to complete on large-scale super-computing infrastructures. An attractive alternative is to use machine learning surrogate models~\cite{miniati,mfluoMLP,mfluoCNN} to achieve computational speed-up. For example, machine learning models can be used to accelerate~\cite{kube2021PICl} or fully replace~\cite{aguilar2021} the field solver block in PIC codes or generate more accurate effective force fields from high-fidelity density functional theory simulations~\cite{Qian2021}.

We adopt a graph-based surrogate model due to their inherent ability to model interactions between mesh elements, which adapts well to structured and unstructured data~\cite{battaglia2018, bronstein2021}. Graph neural networks (GNNs) operate through local message passing~\cite{Scarselli} which enables efficient propagation of information across a computational domain. This work inspired many graph-based networks in the later years. Sanchez-Gonzalez {\it et al.}~\cite{sanchez2020} developed Graph Network-based simulators injecting physics via inductive bias. By combining message passing mechanism and U-Net architecture \cite{ronneberger2015}, Pfaff and colleges developed MeshGraphNet -- a data-driven Graph network that excels for forward mesh-based simulation. In physics, GNNs have been successfully applied in modeling particle-particle~\cite{sanchez2020} and particle-mesh interactions~\cite{pfaff2021} and used in fluid dynamics \cite{pfaff2021, sanchez2020, lam2023}, molecular simulations~\cite{Stocker_2022, Li_2022}, and in kinetic plasma simulations \cite{Carvalho2024, kube2021PICl, aguilar2021, 2025JEl, mlinarević2025particlebasedplasmasimulationusing}. We note that in the above works, the GNN remains purely-data driven --  extracting physics solely from experimental or simulation data.  

For complex problems with uncertain knowledge, physics-informed machine learning (PIML) can combine noisy data with physical constraints to improve accuracy and generalization. Recent work has explored the integration of graph neural networks (GNNs) with physics-informed learning. In Ref.~\cite{seo2019}, Seo and Liu embedded physical equations in GNN for climate forecasting. Reference~\cite{baydin2018} applied PI-GNN to solve spatial-temporal partial differential equations (PDEs) using automatic differentiation due to the high computational complexity of the problem. Ref.~\cite{Chenaud_2024} followed a similar approach; the model managed to predict time evaluation in 2D and 3D physics systems with success. PI-GNN has also been studied with different types of geometries: block-structure grids~\cite{Zou_2024}, structured~\cite{battaglia2018} and unstructured meshes~\cite{Wurth2024}, and uniform grids~\cite{SHI2025}. A physics-constrained Graph Network was used in Ref.~\cite{peng2024} for fast predictions of shock phenomena; however, their training was performed to only learn the time evolution for a single initial condition, and it does not generalize to a range of different initial or boundary conditions. Ref.~\cite{Wurth2024} developed a PI MeshGraphNet architecture based on the finite-element method and used it to learn the problem of smooth heat transport. 

Our work builds upon the aforementioned results and it incorporates a physics informed loss function within the state-of-the-art MeshGraphNet architecture \cite{pfaff2021} to learn the underlying mechanism of shock wave propagation. It uses finite differences to compute spatial derivatives, which has improved accuracy and generalization compared to finite element methods. Our model inherits most of the desirable characteristics from MeshGraphNet while having better performance in generalizing to unseen datasets as well as improved accuracy. Once trained, the model can infer the same problem with various initial conditions without retraining.
 
Learning the dynamics of nonlinear partial differential equations from limited observation data is a central challenge across physics, ranging from shock waves in fluid to plasma dynamics. These systems are governed by Euler conservation laws that generate sharp gradients, posing great difficulties for both numerical solvers and data-driven learning methods. In this study, we use Sedov–Taylor blast waves as a canonical example of strong shocks and self-similar evolution. While recent machine learning approaches have demonstrated success in learning smooth or weakly nonlinear systems, their extension to discontinuous shock dynamics remains limited.

While most of the work mentioned above focus on learning continuous systems, the study of discontinuous shocks with machine learning remains limited, although in recent years it has gained popularity. Ref,~\cite{LiSun2024} and Ref.~\cite{peng2024} used physics-constrained graph network to learn shock propagation with limited or no training data. They succeed in gaining acceleration compared to classical computational fluid dynamics (CFD) methods, but with limited generalization ability to a broader range of different initial conditions.  

Our work shows that embedding physics-informed constraints into a graph neural network fundamentally alters the learning of shock-dominated compressible flows. To avoid the heavy computational cost of automatic differentiation due to complex model structure and large datasets, we employ numerical schemes like the finite-difference method to compute spatial derivatives with sufficient accuracy and efficiency. These constraints act as a physically motivated inductive bias that mitigates overfitting in highly expressive graph models, leading to improved generalization and stability relative to purely data-driven MeshGraphNet architectures. Through controlled comparisons, we demonstrate that physics-informed regularization suppresses spurious oscillations near shocks, improves robustness under sparse observations, and enhances predictive performance on unseen conditions within the trained physical regime.

This paper is structured as follows:  Section II outlines the proposed methodology, including the baseline MeshGraphNet (MGN) model, the design of the physics-informed loss using the Euler equations, and the numerical differentiation schemes. Section III presents the experiments on two hydrodynamic problems involving strong discontinuities (the Sedov-Taylor explosion and the Riemann problem), comparing MGN and physics informed Phy-MGN results for unseen cases. Section IV identifies some of the deficiencies of the current model, difficulties encountered, and possible future improvements. Finally, conclusions are drawn in Section V.

\section{Methodology}

\subsection{Model architecture}
Our model follow a similar architecture to MeshGraphNet (MGN)~\cite{pfaff2021} -- a framework based on PyTorch for learning mesh-based simulations using message-passing Graph Neural Networks and a U-Net architecture~\cite{ronneberger2015}. It consists of Encoder-Processor-Decoder followed by an Euler integrator to process high order dynamics. The encoder layer \texttt{ENCODER} increases the depth of input feature vector thereby capturing the abstract representations of the input. The \texttt{PROCESSOR} uses a Message Passing mechanism, which allows information to propagate via node connections and capture long-range dependencies. The decoder layer \texttt{DECODER} reduces the depth of feature maps and reconstruct predicted dynamics from high-dimensional features. Given the graph state at time \(t\), the model is trained to predict the dynamics of a given state at \(t\) or directly predict the states in the next time step \(t+\Delta t\) where \(\Delta t \) is the interval of the time step and remains constant for all input graphs \(\mathcal{G}^t\).

\subsubsection{Graph structure} 
\begin{figure*}
    \centering
    \includegraphics[width=\textwidth]{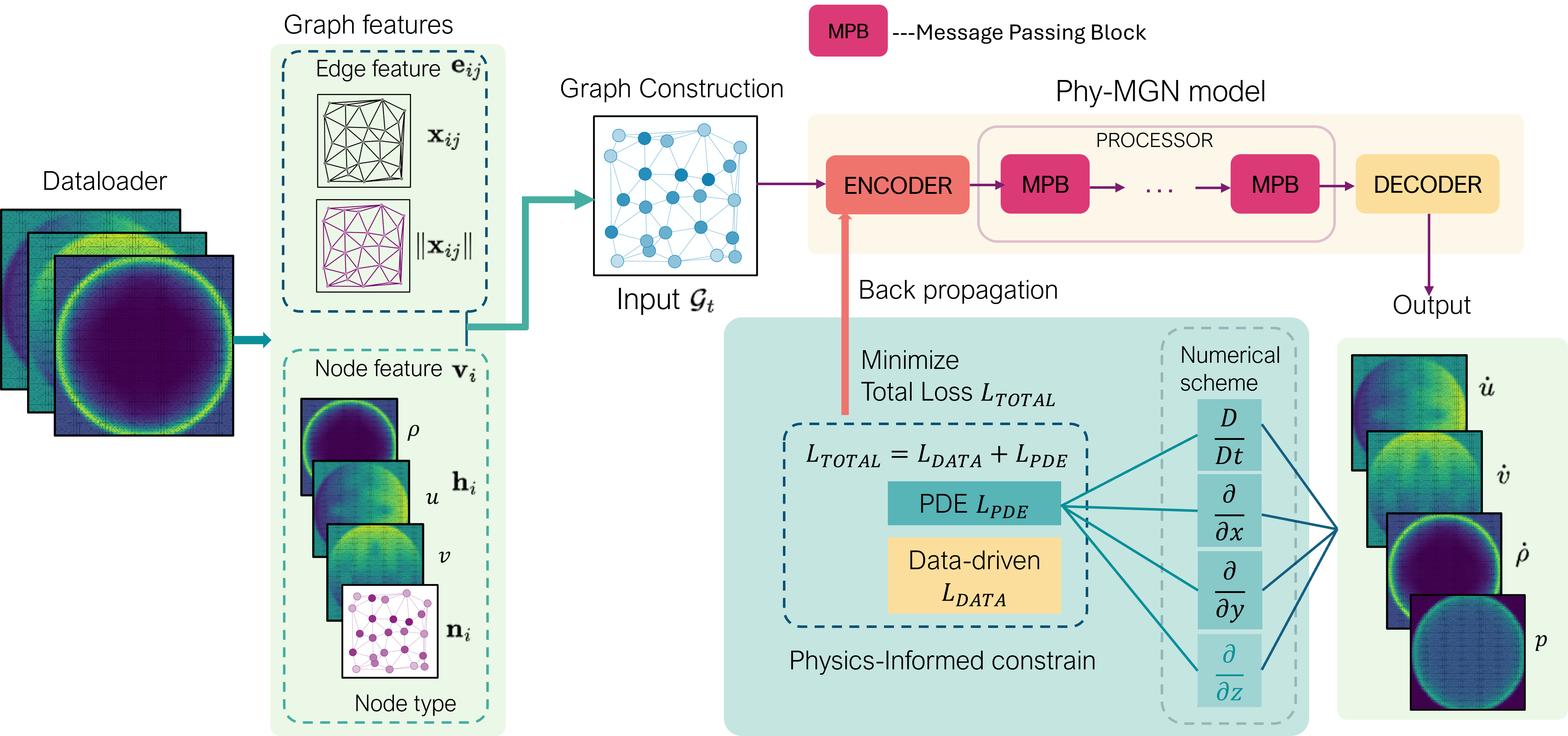}
    \caption{\textbf{Diagram of Phy-MGN training process}. The dataloader loads a batch of graph data into the trainer, where the node and edge features are concatenated and reconstructed into a graph object before being passed into the model. The model performs a forward pass and generates outputs that represent both the state dynamics and predicted future states. By combining these outputs with the input states, the complete state at the next time step is obtained, which can then be used within the governing PDE to compute spatial derivatives. The PDE residual, derived from these computations, serves as the physics-informed loss. Together with the data-driven loss, it forms the total loss function. This total loss is then backpropagated through the network to update the model parameters. The entire process is repeated iteratively until the training converges.}
    \vspace{1em}
    \label{fig:PI-MGN}
    
\end{figure*}

The training data in this paper is generated in multi-physics simulation code FLASH \cite{Fryxell2000}. For simplicity, we used a uniform static grid for all of our simulations. In Fig.~\ref{fig:PI-MGN}, consider a multi-dimensional grid (1D, 2D or 3D) as domain \(\Omega\) with open boundary, denoted by \(\mathcal{G} = (\mathcal{N}, \mathcal{E})\), with grid nodes \(\mathcal{N}\) connected by edge \(\mathcal{E}\). There are \(\mathrm{N}\) nodes in a graph where each node \(v_i\) is associated with a unique coordinate \(\mathbf{x_i}\) and a feature vector \(\mathbf{v_i}\) that contains system state feature vector \(\mathbf{h_i}\) and one-hot encoded node type vector \(\mathbf{n_i}\)  that distinguish dynamic nodes from boundary nodes.  Each node is connected by bidirectional edges \(e_{ij}\) from sender node \(v_i\) to receiver node \(v_j\). Like nodes, each edge is associated with an edge feature \(\textbf{e}_{ij}\) that contains the relative displacement vector between two adjacent node \(v_i\) and \(v_j\) : \(\mathbf{x_{ij}} = \mathbf{x_i - x_j}\) and distance \(\Arrowvert \mathbf{x_{ij}} \Arrowvert\). The edge features are time-independent for static grids. Detals of model hyperparameters can be found in Appendix~\ref{app:B}.

\subsubsection{Encoder} 

The encoder transforms input grid into a high-level latent graph composed of nodes and edges via a MLP (Multi-Layer Perceptron)~\cite{MURTAGH1991, Popescu2009}. The grid vertices are represented as nodes, and the connections between vertices in the input grid space are represented as bidirectional edges. 
The \texttt{ENCODER} layer consists of two encoders for nodes, \texttt{ENCODER}$^{\mathcal{N}}$, and for edges,  
\texttt{ENCODER}$^{\mathcal{E}}$, respectively. The encoder output for the grid at time step \(t\) is a latent graph with transformed node and edge features 
\begin{eqnarray}
    \mathcal{G}^t = \{ \texttt{ENCODER}^{\mathcal{N}}(\{ \textbf{v}_i\}^N_{i=1}), 
    \texttt{ENCODER}^{\mathcal{E}}(\{\textbf{e}_{ij}\}^N_{i, j = 1})\},
\end{eqnarray}
with node feature \(\textbf{v}_i = [ \textbf{h}_i, \textbf{n}_i]\), and edge feature \(\textbf{e}_{ij} =[\mathbf{x_{ij}}, \Arrowvert \mathbf{x_{ij}} \Arrowvert] \).

\subsubsection{Processor}

\begin{figure}[b]
    
    \includegraphics[width=0.45\textwidth]{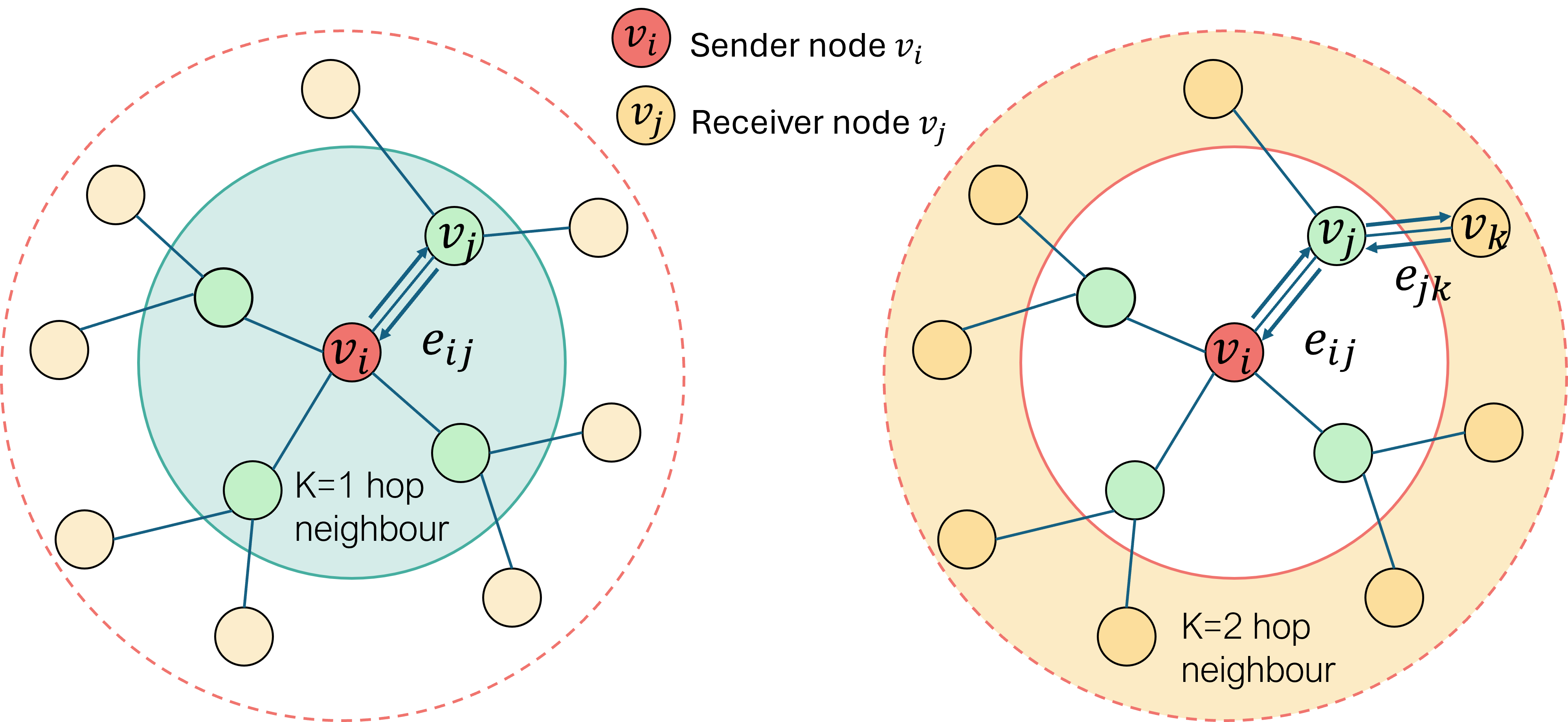}
    \caption{\textbf{Diagram of k-hop neighborhood}. When message passing is applied k-times (k=2 here), each node aggregates information from all nodes within its k-hop neighborhood~\cite{k-hopping}.}
    \vspace{1em}
    \label{fig:k-hop}
    
\end{figure}

The processor deals with message-passing. Message passing in graph neural network is an iterative process that update the embedding on each node by aggregating information from neighboring nodes and incident edges. This process enables neural network to capture dependencies and interactions among graph nodes. MGN contains several message passing blocks (MPBs), each represents a message passing iteration. A message passing block has two update functions, \texttt{UPDATE}$^\mathcal{N}$ and \texttt{UPDATE}$^\mathcal{E}$, for nodes and edges respectively, as well as an aggregation, \texttt{AGGREGATE}$^\mathcal{E}$, a sum operation in our case,  followed by a layer normalization step \cite{ba2016}, with a residual connection \cite{he2015}.The mathematical operation when passing through an MPB can be expressed as:

\begin{equation}
\textbf{e}^{'}_{ij} = \texttt{UPDATE}^\mathcal{E}(\textbf{e}_{ij}, \textbf{v}_i, \textbf{v}_j),
\end{equation}
\begin{equation}
    \bar{\textbf{e}}_i^{'} = \texttt{AGGREGATE}^\mathcal{E}(\{ \textbf{e}_{ij}^{'}\}^m_{j=1}),
\end{equation}
\begin{equation}
    \textbf{v}_i^{'} = \texttt{UPDATE}^{\mathcal{N}}(\bar{\textbf{e}}_i^{'}, \textbf{v}_i),
\end{equation}
where \(\textbf{v}_i\) and \(\textbf{v}_j\) are the encoded latent vector of sender node and receiver node, \(m\) is the number of neighboring node \(v_j\) around \(v_i\). After passing through \(k\) message passing blocks (see Fig.~\ref{fig:k-hop}), the message would be propagated to the $k$-hop neighbor of node \(v_i\), simulating the range of interaction a node has to its surroundings. This mechanism is shown in detail in Fig.~\ref{fig:k-hop}. 

\subsubsection{Decoder}

\begin{figure*}
\includegraphics[width=\textwidth]{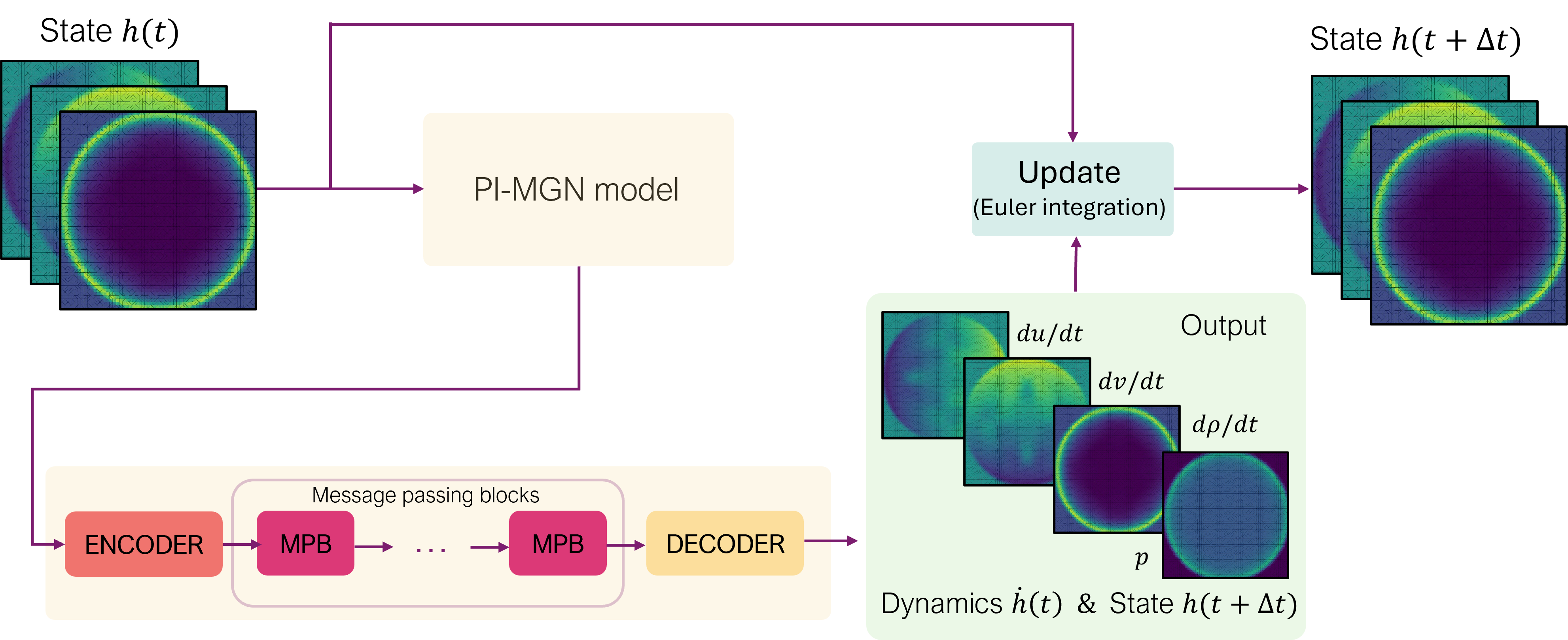}
\caption{\textbf{Diagram of Phy-MGN inference process}. Phy-MGN consists of an Encoder-Processor-Decoder architecture. The encoder transforms input graph $\mathcal{G}_t$ into latent space, the processor performs several rounds of message passing, the decoder then computes the dynamics and future states to update input state, producing states and graph in the next timestep $\mathcal{G}_{t+1}$. This process is applied iteratively to produce predictions along inference time.}
\label{fig:MGN_inference}
\end{figure*}

Similar to \texttt{ENCODER}, a \texttt{DECODER}
is an MLP with the same structure as \texttt{ENCODER}. It consists of two decoders: for nodes, \texttt{DECODER}$^{\mathcal{N}}$, and for edges, \texttt{DECODER}$^{\mathcal{E}}$, respectively. It converts the information from a high-dimensional latent vector to the dimensions of the output features. This decoding process can be expressed as 
\begin{eqnarray}
    \mathcal{V}^t_{out} = \{ \texttt{DECODER}^{\mathcal{N}}(\{ \textbf{v}^{'}_i\}^N_{i=1}),
    \texttt{DECODER}^{\mathcal{E}}(\{\textbf{e}^{'}_{ij}\}^N_{i, j = 1})\}.
\end{eqnarray}
The output features,  \(\mathcal{V}^t_{out}\), of \texttt{DECODER} are interpreted as a mixture of state dynamics at \(t\) and future states at \(t+\Delta t\). State dynamics refers to the per-timestep change of the state. We use inference to  predict future velocities \(v_i^{t+\Delta t}\). This is done by performing a forward-Euler integration
\begin{equation}
    v_i^{t+\Delta t} = v_i^{t} + \Delta \textbf{v}_i^{t},
\end{equation}
as illustrated in Fig.~\ref{fig:MGN_inference}.
Additional output features may also include the pressure ($p$) of the fluid as a state variable. Since this is a zeroth-order quantity (in time), it does not require any further integration and it is given as a future state as \(p(t+\Delta t)\) in \(\mathcal{V}^t_{out}\).  

\subsection{Conservation equations}
In our numerical experiments, the governing equations are Euler equations. These express conservation laws of mass, momentum and energy. 
These are specialized for an inviscid, compressible gas in Cartesian coordinates:
\begin{equation}
\label{eq:Euler}
    \frac{\partial U}{\partial t} + \nabla  \cdot  f(U) = 0, \ x\in \Omega \subset \mathbb{R}^d, \ d = 1, 2, 3, \ t\in (0, T], 
\end{equation}
where, in 2-dimensions, 
\begin{eqnarray}
&&U=
    \begin{pmatrix}
        \rho \\
        \rho u_1 \\
        \rho u_2 \\
        \rho E
    \end{pmatrix}, \ f = (G_1, G_2), \nonumber
\end{eqnarray}
with,
\begin{eqnarray}
    G_i(U) = 
    \begin{pmatrix}
        \rho u_i \\
        \delta_{i1}p + \rho u_1u_i \\
        \delta_{i2}p + \rho u_2u_i \\
        pu_i + \rho u_i E
    \end{pmatrix},\ i=1, 2,
    \label{Euler}
\end{eqnarray}
where \(\rho\) is density, \(p\) is pressure, \(u_1\) and \(u_2\) are velocities in two orthogonal directions (\(x_1\) and \(x_2\)), $E$ is the total energy and \(\delta_{ij}\) is Kronecker delta. The total energy is the sum of internal energy, \(\epsilon\), and kinetic energy per unit mass:
\begin{equation}
    E = \frac{1}{2}|\textbf{u}|^2 +\epsilon,
\end{equation}
and, in the case of ideal gas, the internal energy can be expressed as
\begin{equation}
    \epsilon = \frac{p}{(\gamma - 1)\rho}
\end{equation}
where \(\gamma\) is the ratio of specific heats (\(\gamma = 1.4 \) is used for all simulations below). 
As the energy equation is not directly solved in the simulations, we only used mass and momentum conservation in order to enforce the physics-constraint loss function.



\subsection{Computing derivatives}
When computing derivatives in a physics informed neural network (PINN), the most common approach is automatic differentiation~\cite{baydin2018}. Auto‐differentiation‐based PINNs embed the PDE directly within a neural network by treating the solution \(U(\textbf{x}, t)\) as a continuous function of coordinates and time. 
It requires exact spatio-temporal coordinates at inference, which is incompatible with autoregressive time-series models like ours that operate only on successive states rather than absolute coordinates and time steps. Moreover, automatic differentiation must retain the entire computational graph so that gradients can be propagated backwards through every operation -- all intermediate values in the forward pass must be cached to perform the backward pass by the chain rule. In graph-based models with large amount of training data, this this can lead to very high memory usage and low training speed, making auto-differentiation PINNs computationally expensive compared to numerical-based differentiation methods. As our training data is based on a structured Cartesian grid, we opted for a finite difference method to compute PDE residuals.

\subsubsection{Spatial derivative}

We use a finite difference method to calculate spatial derivative due to its simplicity. Assuming a uniform Cartesian grids with a uniform mesh of size \(\Delta x\), we use the central difference formula, which is second-order accurate, for the first and second derivatives:
\begin{equation}
    \frac{\partial}{\partial x}f(x) = \frac{f(x+\Delta x) - f(x - \Delta x)}{2\Delta x},
\end{equation}
\begin{equation}
    \frac{\partial^2}{\partial x^2}f(x) = \frac{f(x+\Delta x) - 2f(x) + f(x - \Delta x)}{\Delta x^2}.
\end{equation}

\subsubsection{Time derivative}

As for any time‐series learning, the model predicts the solution at the next time step \(t+\Delta t\) based on the states at the current time \(t\). By subtracting the states across two consecutive time steps, one can obtain an estimate of the first‐order temporal derivative. Two common numerical schemes exist for this purpose: the forward Euler (explicit) method and the backward Euler (implicit) method. The forward Euler method is fully explicit and evaluates the spatial derivatives at the known time step \(t\), resulting in the following discretizations of Euler equation (\ref{eq:Euler}):
\begin{equation}
\frac{U(t+\Delta t) - U(t)}{\Delta t} + \sum_i \frac{\partial f(t)}{\partial x_i} = 0.
\end{equation}
Since every state has to be the inferred output of the model, the model’s predictions need to be rolled forward autoregressively: the predicted \(U(t+\Delta t)\) is fed back into the network to produce \(U(t+2\Delta t)\), so that the forward Euler discretization becomes
\begin{equation}
\frac{U(t+2\Delta t) - U(t+\Delta t)}{\Delta t} + \sum_i \frac{\partial f(t+\Delta t)}{\partial x_i} = 0.
\end{equation}
In comparison, implicit or semi-implicit methods like backward Euler method provide
some stability for the outputs in admissible time steps. We attempted two ways to compute PDE: 1) forward Euler method and 2) backward Euler method which provides simplicity and stability. 

\subsubsection{Modification of the PDE using material derivatives}

The implicit method approximates the time derivative using
\begin{equation}
    \frac{dU(t)}{dt} \approx \frac{U(t+\Delta t) - U(t)}{\Delta t},
\end{equation}
which inherently depends on the future state \(U(t+\Delta t)\) and thus tends to incorporate the influence of surrounding and downstream regions. Physically, this corresponds to approximating the material derivative rather than merely the partial time derivative. For a macroscopic field \(U(\mathbf{x},t)\), the material derivative is defined as
\begin{equation}
    \frac{D}{Dt}U = \frac{\partial U}{\partial t} + \mathbf{u} \cdot \nabla U,
\end{equation}
where \(\mathbf{u} = d\mathbf{x}/dt\) denotes the flow velocity. In contrast to the partial derivative \(\partial U/\partial t\), which measures only the local change at a fixed point in space, the material derivative additionally accounts for the convection of information carried by the flow field. The difference between these two derivatives becomes particularly significant in regimes with strong spatial gradients, such as near shock fronts or rapidly propagating waves, where downstream and neighboring effects strongly influence the evolution of the solution.

We modify the governing PDE (Eq.~\ref{eq:Euler}) to account for this effect. They are written explicitly as:
\begin{equation}
    \frac{D \rho}{D t} + \rho \Big(\frac{\partial u_1}{\partial x_1} + \frac{\partial u_2}{\partial x_2} \Big )= 0,
\end{equation}
\begin{equation}
    \frac{D}{D t}(\rho u_1) + \frac{\partial p }{\partial x_1} + \rho u_1 \frac{\partial u_1}{\partial x_1} +\frac{\partial}{\partial x_2}(\rho u_1u_2) = 0,
\end{equation}
\begin{equation}
    \frac{D}{D t}(\rho u_2) + \frac{\partial p }{\partial x_2} + \rho u_2 \frac{\partial u_2}{\partial x_2} +\frac{\partial}{\partial x_1}(\rho u_1u_2) = 0.
\end{equation}
These equations will be used as the governing PDE for the physics-informed loss.

\subsection{PDE-based loss}
One common way to implement physics constraints in any ML model is to use a physics-informed loss function. For Phy-MGN, the underlying principle is for the model to learn from data and physics constraints so that the network can fit the observations while satisfying the physics described by partial differential equations. The loss function combines both data-driven loss, \(\mathcal{L}_{DATA}\)m and physics-informed loss, \(\mathcal{L}_{PDE}\). The total loss function \(\mathcal{L}_{total}\)is therefore given by 
\begin{equation}
    \mathcal{L}_{total}(\theta; \mathcal{V}^t_{in}) = \mathcal{L}_{DATA}(\theta; \mathcal{V}^t_{in}) + \sum_i^N\lambda_{i}\mathcal{L}^{i}_{PDE}(\theta; \mathcal{V}^t_{in}),
\end{equation}
where \(\theta\) represents the set of trainable parameters in the model, \(\mathcal{V}^t_{in}\) is the set of input node features at timestep \(t\) and $N$ is the number of conservation equations used in physics-informed loss.
For inviscid, compressible flow governed by Euler equation, there are several states of interest included in the input node features: velocity \(\textbf{u}(\textbf{x}, t)\), density \(\rho(\textbf{x}, t)\), and pressure \(p(\textbf{x}, t)\). With these variables, we choose Eq.~\ref{eq:Euler} as the governing PDE and use a finite difference method to calculate the spatial derivatives. The PDE residual is given by 
\begin{equation}
    \mathcal{R}(\mathcal{V}^t_{out}) = \frac{D}{D t}U(\mathcal{V}^t_{out}) + \mathcal{F}(\mathcal{V}^t_{out}, \nabla_{\textbf{x}_1}\mathcal{V}^t_{out}, \nabla_{\textbf{x}_2}\mathcal{V}^t_{out}),
\end{equation}
where 
\begin{equation}
    \mathcal{V}^t_{out} =  \mathrm{PhyMGN}(\theta;\mathcal{V}^t_{in}), 
\end{equation}
is the model outputs given trainable parameters \(\theta\) and input node feature \(\mathcal{V}_{in}^t\). The model outputs includes velocity, density and pressure.

Unlike smooth, continuous fluids where the solution is differentiable everywhere, shock-dominated problems exhibit strong discontinuities which invalidates Euler equation that assumes smooth and differentiable flow, leading to high residual values near the shock front in the simulation data. Because of these numerical errors in the ground truth, we are not able to only adopt a data-free approach in the loss function, but we have to combine the physics-informed loss with a data-driven loss. In such cases, minimizing the PDE residual to near-zero is inappropriate, as it would contradict the observed (ground truth) data. To address this, we first compute the residual map from the observational data before backpropagation and subtract it from the model-predicted PDE residual. This approach shifts the training objective to minimizing the difference between the model-inferred residual and the ground-truth residual. We are learning consistency with the conservation defect of the solver, rather than strictly enforcing exact conservation. As a result, the data-driven loss and the physics-informed loss become aligned rather than conflicting, improving training stability and physical consistency. For each conservation equation, The PDE residual of the observation data \(\mathcal{R}_{obs}\) is given by
\begin{equation}
    \mathcal{R}_{obs}(y^t) = \frac{D}{D t}U(y^t) + \mathcal{F}(y^t, \nabla_{\textbf{x}_1}y^t, \nabla_{\textbf{x}_2}y^t),
\end{equation}
where \(y^t\) is the observation data. The physics-informed loss for the \(i_{th}\) conservation equation is then defined by 
\begin{equation}
    \mathcal{L}^i_{PDE}(\theta; \mathcal{V}^t_{in}, y^t) = \parallel \mathcal{R}^i(\mathcal{V}^t_{out}) - \mathcal{R}^i_{obs}(y^t) \parallel ^2,
\end{equation}
where \(\parallel \cdot \parallel\) represents the \(L2\) form. The data driven loss, \(\mathcal{L}_{DATA}\), is the mean square error (MSE) loss between the predicted value and the observation data: 
\begin{equation}
    \mathcal{L}_{DATA}(\theta;\mathcal{V}^t_{in}, y^t) = \parallel \mathcal{V}^t_{out} - y^t\parallel^2.
\end{equation}
Therefore, the total loss function is described by
\begin{eqnarray}
    &&\mathcal{L}_{total}(\theta;\mathcal{V}^t_{in}, y^t) \\  &&= \parallel \mathcal{V}^t_{out} - y^t\parallel^2\nonumber + \sum_i^N\lambda_{i}\parallel \mathcal{R}^i(\mathcal{V}^t_{out}) - \mathcal{R}^i_{obs}(y^t) \parallel ^2,
\end{eqnarray}
where the weight factor \(\lambda\) is chosen so that the physics-informed loss is approximately \(10\%\) of data-driven loss, or 
\begin{equation}
     \sum_i^N\lambda_{i}\mathcal{L}^i_{PDE} \sim 0.1 \ \mathcal{L}_{DATA}.
\end{equation}
The proper choice of \(\lambda_i\) ensures that the gradient contribution from the data-driven term, \(\mathcal{L}_{DATA}\), remains dominant while the physics-informed term, \(\mathcal{L}_{PDE}^i\), continues to provide a non-negligible supportive role during training. In practice, this choice is important because the PDE residual is affected by numerical errors and noise arising from the finite-difference discretisations and numerical noise in simulation data (see Appendix \ref{app:E}). Consequently, \(\mathcal{L}_{PDE}^i\) cannot perfectly describe the underlying system dynamics, making it preferable for \(\mathcal{L}_{DATA}\)to have a stronger influence while \(\mathcal{L}_{PDE}^i\) gives additional regularization based on the physical equations.

\section{Numerical experiments}

In this section, we demonstrate the capability of Phy-MGN in learning shock propagation speed in the Sedov-Taylor explosion problem, which exhibits strong discontinuities. As ground truth, the training sets, validation sets and test sets are generated from FLASH, a multi-physics application code that is capable of simulating hydrodynamic and magneto-hydrodynamic simulations with high fidelity. Dataset configurations are provided in Appendix \ref{app:A}. The code is written using Pytorch, developed and adapted based on the MeshGraphNet code base. 

\subsection{Sedov-Taylor problem}
The Sedov explosion models the self-similar evolution of a spherical shock wave generated by a near-delta-function pressure perturbation. 
The simulation is done for the ideal hydrodynamics case -- no convection or diffusion processes is included. The simulation uses a 2nd-order MUSCL scheme for data reconstruction, and a Riemann solver for flux calculation. In this setup, a fixed amount of energy $E$ is deposited within a small initial radius $\delta r$, producing a radially propagating shock in a homogeneous medium of ambient density $\rho_A$. The shock front travels at supersonic velocity and consists of highly compressed material, while the ambient pressure remains fixed at $p = 1\times10^{-6}$ Pa , except at the explosion centre where the perturbation is applied \cite{flash_user_guide_section}. By varying the ambient density, the model can be trained to learn the propagation of the shock front given an initial state. For training, we select 24 densities, 9 of them in the low density regime (below solid densities) \(\rho_A \in [0.1, 0.9]\) g/cm$^3$, and the other 15 in the high density regime \(\rho_A \in [1, 15]\) g/cm$^3$. More details of the dataset can be found in Appendix~\ref{app:A}.

\begin{figure*}
\includegraphics[width=\textwidth]{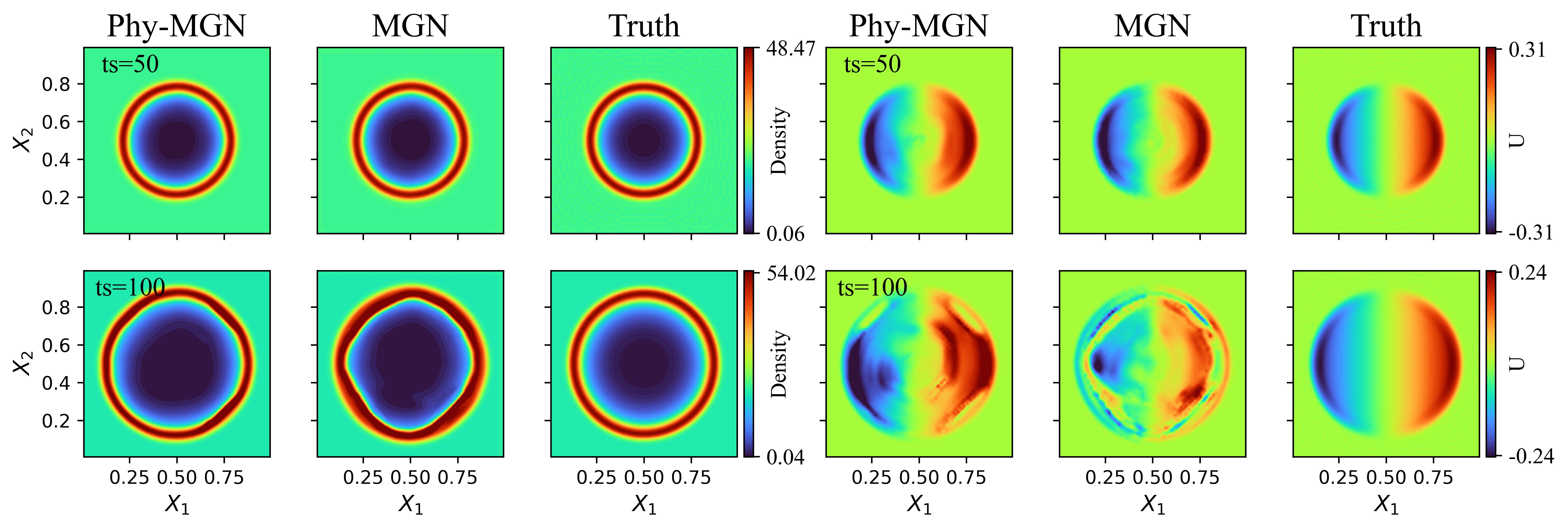}

\caption{\label{fig:sedov_pred} \textbf{Model predictions on an unseen test case}. The figure above presents Phy-MGN and MGN predictions of density and x-velocity (U) for a sample with ambient density 19, a test case outside the training parameter space, shown at the 50th (top row) and 100th (bottom row) inference time steps. The model was trained on data processed with a smoothing algorithm. The comparison shows that both models interpolate reliably early in the rollout. By the 100th step, MGN accumulates noticeable errors, with a more distorted shock front and noisier velocity fields—while Phy-MGN prediction maintains a clearer ring structure and more coherent flow, showing improved stability at longer rollout times.}
\end{figure*}

Incorporating a physics-informed loss term alongside the data-driven loss substantially improves the model’s generalization capability, enabling it to perform reliably even in scenarios that moderately fall outside the training distribution. This property is demonstrated in Fig.~\ref{fig:sedov_pred} , which compares the inference results of Phy-MGN, baseline MGN, and the FLASH reference solution for ambient density of $\rho_A=19$ g/cm$^3$, which lies outside the training dataset. The inference uses the 20$th$ timestep in the simulation as the initial state and does a rollout for another 100 iterations. At the 50$th$ timestep, both Phy-MGN and MGN produce accurate predictions that align with the ground truth, but with Phy-MGN having a higher accuracy. Each timestep \(\Delta t\) elapses 0.005s. The predictions gets worse in later timesteps as error accumulates: Phy-MGN is still able to capture the shock front accurately against error accumulation while MGN result shows a higher divergence from FLASH simulation prediction (the ground truth). By embedding physical constraints into the training process, the physics-informed loss prevents the model from overfitting too closely to training dataset. Instead, it steers the model toward solutions aligned with the governing equations, thereby enhancing robustness and extending predictive power to unseen conditions. A more thorough assessment of the Phy-MGN performance can be found in Appendix~\ref{app:C} and \ref{app:D}.

\subsection{Runtime performance}
In Table \ref{tab:computation time} we compare the computation time of Phy-MGN and the FLASH code for the Sedov–Taylor explosion at ambient densities greater than \(10 g/cm^3\). For lower ambient densities, FLASH simulations typically require longer runtime. Phy-MGN inference results are benchmarked on a server-class GPU (NVIDIA Tesla T4), and a server-class CPU with 32 cores (AMD EPYC 7502P CPU @ 2.5GHz), while FLASH is executed on the same CPU node for consistency.

\begin{table}[b]
\caption{\label{tab:computation time}%
The prediction time cost of FLASH and Phy-MGN model for Sedov Taylor problem.}
\begin{ruledtabular}
\begin{tabular}{ccccc}
Cases&size&
\multicolumn{1}{c}{\textrm{FLASH time (s)}}&
\multicolumn{1}{c}{\textrm{Phy-MGN time (s)}}\\
\hline
Sedov& \(64 \times 64\) & 122.09\footnote{Could be accelerated to around 31.8 s by setting lower cfl condition and using more CPU cores}&1.16\\

\end{tabular}
\end{ruledtabular}
\end{table}

During training, one major concern is the computational time of the PDE residual. With optimised programming, each training iteration—including backpropagation—takes approximately 0.188 s for purely data-driven training. When the PDE residual is included, the iteration time increases slightly to an average of 0.194 s for finite-difference schemes. This modest overhead reflects the additional computations required by the physics-informed loss as expected.  All experiments are trained on a high-performance computing (HPC) system using 8 NVIDIA A100 GPUs, with each training run requiring approximately 24 hours to complete with 400 epochs.

\section{Discussion and conclusion}

In this work, we have explored the effect of physics-informed constraints in controlling the generalization behaviour of a graph neural network for shock-dominated compressible flows. By augmenting the baseline MeshGraphNet architecture with a finite-difference-based physics-informed loss derived from the Euler equations, we demonstrated that by incorporating the physical structure of the equations as a soft constraint, reduces overfitting and improves stability when learning two-dimensional discontinuities from augmented simulation data. While the presence of numerical noise near shock fronts prevents fully data-free training, the physics-informed formulation consistently yields improved predictive performance and robustness relative to purely data-driven models.

From a learning perspective, the addition of physics-informed constraints acts as a physically motivated inductive bias that shifts the model away from spurious correlations in the training distribution toward representations that better respect conservation laws. Although the pure MGN is learning the solver-specific correlations, Phy-MGN constrains the representation towards conservation-consistent manifold. This effect manifests most clearly in the enhanced generalisation to unseen regimes, particularly for slower-propagating shocks, while maintaining competitive accuracy within the training distribution. Our results also highlight the strong representational capacity of the baseline MeshGraphNet for highly nonlinear dynamics and clarify how physics-based training can extend this capability to a broader parameter space.

Several limitations of the present study merit discussion. First, the physics-informed loss contributes a relatively small fraction of the total training objective. This choice reflects two fundamental constraints: (i) the presence of irreducible numerical noise in the ground-truth data near shock fronts, as discussed in the Appendix~\ref{app:E}, which limits the reliability of PDE residuals; and (ii) the finite spatial resolution of the training data, which restricts the accuracy with which sharp gradients and discontinuities can be resolved. This can be improved by using higher resolution data with more stable numerical schemes. As a result, the physics-informed term alone cannot dominate the optimization process and the overall performance of Phy-MGN remains primarily driven by the data-fitting component. Nevertheless, despite its limited weight, the physics-informed loss consistently improves generalisation and stability.



Overall, Phy-MGN has inherited the strong learning ability of the original MGN and has been shown to surpass it in terms of accuracy and generalisation. Being fully differentiable in parameter space, this framework could be applied to parameter scanning and inverse design optimization \cite{Kasim2019}, where a large amount of simulations has to be done and would save a great amount of time. 
Additionally, this framework can be naturally extended to three-dimensional simulations with a significant increase in domain size, or to explore problems with richer physical processes such as diffusion, turbulence, and magnetic fields, ultimately advancing the role of machine learning in scientific computing for extreme physical phenomena. 

\begin{acknowledgments}
The authors are grateful for the computing resources provided by the Science and Technology Facilities Council (STFC) Scientific Computing Department’s SCARF cluster. This research is funded by European Space Agency (Discovery program, ESA Contract No. 4000141210/23/NL/GLC/my). The authors thank the European Space Agency and STFC for its financial support.
\end{acknowledgments}

\appendix

\section{Model details}
\label{app:B}
\subsection{ENCODER \& DECODER}

For nodes and edges, the concatenated features above are then transformed into a latent vector of size 128 (which is the optimized value determined through empirical methods), through a Multi-Layer Perceptron (MLP) \cite{MURTAGH1991, Popescu2009}, a simple feed-forward neural network. In MGN, the MLP for \texttt{ENCODER}, \texttt{PROCESSOR} and \texttt{DECODER} consists of two hidden layers with an hidden dimensionality of 128, using activation functions \(\mathrm{ReLU}\) \cite{agarap2019deeplearningusingrectified} between layers.

\subsection{PROCESSOR}

 The processor in MGN contains 12 message passing blocks. Each block consists of a message-passing layer with a residual connection \cite{he2015} followed by a layer normalisation step \cite{ba2016}. Residual connection adds the input of a layer to its output, letting information bypass the layer and preventing it diminishing to zero, thereby helps address the vanishing gradient problem and allows training to be tractable .  Layer Normalisation normalizes the activations in a neural network by calculating the mean and standard deviation for each individual layer. These statistics are then used to scale and shift the activation values ( to between 0 and 1 ) to stabilize training and improve gradient flow. It also works well with data-parallelism. The model hyperparameter are listed in Table \ref{tab:table5}.
\begingroup
\setlength{\tabcolsep}{80pt}
\begin{table*}[htbp]
\caption{\label{tab:table5} Hyperparameter configuration of the MeshGraphNet model used for learning Sedov problem.}
\begin{ruledtabular}
\begin{tabular}{cc}

 \emph{Hyperparameter} & Sedov  \\ \hline
Layers in node encoder MLP   & 2  \\
 Layers in edge encoder MLP & 2  \\
 Layers in node decoder MLP   & 2  \\
 Number of processor layers & 12 \\
 Layers in node processor & 2  \\
 Layers in edge processor & 2 \\
 Neurons per layer & 128  \\
 Normalisation type & Layer Normalisation  \\
 activation function & ReLU\\
 connection type & Residual connection \\
 epoch to add PDELoss & 50 \\
 optimizer & AdamW \\
 learning rate strategy & Exponantial Decay \\
 Initial learning rate & 1e-3 \\

\end{tabular}
\end{ruledtabular}
\end{table*}
\endgroup

\section{Dataset details}
\label{app:A}
Appendix A lists the details for all of our datasets above. All data are generated in multi-physics code FLASH. HD simulations Sedov-Taylor explosion and Relativistic Riemann problem are computed using the Hydro module in FLASH. Due to FLASH architecture, all FLASH generated simulation are based on uniform grid. We disabled adapted mesh refinement so we train on stationary grid. For HD simulations, we include density, pressure, x-velocity, y-velocity, and temperature as physical states and cell structures (node-node connection and node coordinates). The time it takes for FLASH to run a single simulation varies depends on the initial parameters. For Sedov explosion for example, one single simulation can take from 20s to 120s.

\begin{table*}
\caption{\label{tab:table2} Parameters of the dataset used for training}
\begin{ruledtabular}
\begin{tabular}{cccccccc}

 \emph{Dataset}  & \emph{Grid type} & \emph{Domain size (cm)} & \emph{$\Delta x$ (cm)} & \emph{$\#$ steps} & \emph{$\Delta t$ (s)} & \emph{$\#$ samples }& \emph{variable}  \\ \hline
 Sedov      & 2D square & $1.0\times 1.0$ & 1/64 & 120  & 0.005  & 500 & $\rho_0\footnote{Ambient density}, \textbf{x}_0$\footnote{Coordinate of explosion centre.} \\
 
\end{tabular}
\end{ruledtabular}
\end{table*}

And for graph construction, we concatonate node feature and edge feature to construct an input graph for training. Table.~\ref{tab:table3} shows the input features and output features for the Sedov datasets. 

\begin{table*}
\caption{\label{tab:table3} shows the input features of nodes and edges and output node features for the datasets respectively.}
\begin{ruledtabular}
\begin{tabular}{ccccc}

 \emph{Dataset} & \emph{System} & \emph{inputs $\textbf{e}_{ij}$} & \emph{inputs $\textbf{v}_i$} & \emph{outputs}\\ \hline
 Sedov    & Hydrodynamic & $\textbf{x}_{ij}, \Arrowvert \mathbf{x_{ij}} \Arrowvert$  & $\textbf{u}_i(t), \textbf{v}_i(t), \rho_i(t), \textbf{n}_i$ & $\dot{\textbf{u}}_i(t), \dot{\textbf{v}}_i(t), \dot{\rho}_i(t),  p_i(t+\Delta t)$ \\
 
\end{tabular}
\end{ruledtabular}
\end{table*}

\subsection{Smoothing algorithm for Sedov problem}

Because FLASH relies on finite-volume numerical schemes, different solvers and reconstruction methods introduce different forms of numerical anisotropy. These scheme-dependent instabilities often manifest as subtle but nonphysical patterns—such as grid-aligned structures or directional artifacts—that contaminate the simulation results. Such artifacts are especially problematic when the data are used for machine-learning training, since the model may learn these numerical patterns instead of the true physical behavior. We found that the second-order MUSCL scheme in FLASH produces simulations with the lowest level of numerical anisotropy.  

To address the problem above, we apply a simple but effective post-processing procedure: rotational averaging about the explosion center for every sample in the dataset. By averaging the solution over multiple angular orientations, this method suppresses grid-aligned numerical noise while preserving the underlying radial physics of the explosion.This is a way to introduce symmetry bias to improve prediction~\cite{Fuiza}.

\section{A comprehensive comparison of Phy-MGN and MGN}
\label{app:C}
\begin{figure*} [htb!]
\includegraphics[width=\textwidth]{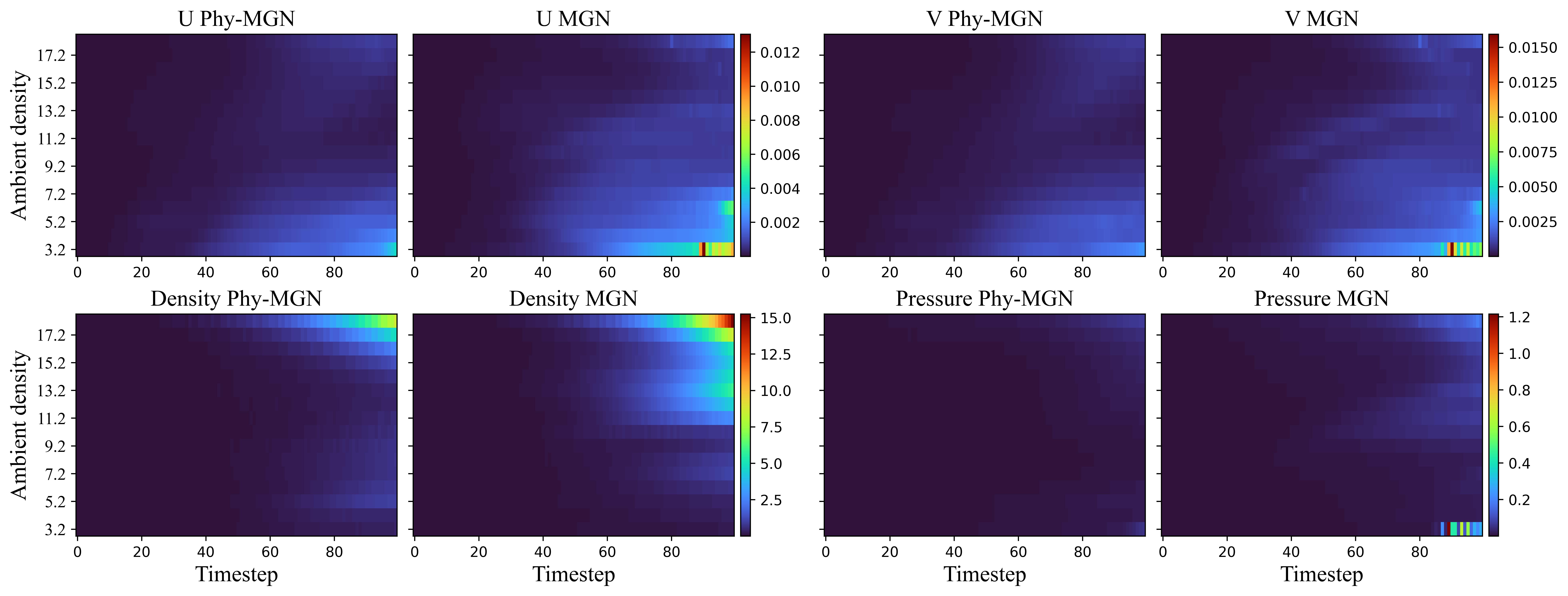}

\caption{\label{fig:Phy-MGN_vs_MGN} \textbf{The mean squared error (MSE) of the Sedov problem} is plotted for both Phy-MGN and MGN across all four state variables: density, pressure, x-velocity U, and y-velocity V. The models are trained using smoothed data. The resulting colormap indicates that Phy-MGN maintains lower errors over a longer prediction horizon than MGN, demonstrating greater robustness to error accumulation during iterative rollout, and improvement in long term stability.  }
\end{figure*}

In this section, we present a more comprehensive assessment of Phy-MGN and MGN performance, trained with smoothed data. Table.~\ref{tab:quantitive} presents a quantitive evaluation showing the mean squared error of Phy-MGN and MGN over all 16 test cases. Phy-MGN always yields the lowest error. The Details of Phy-MGN and MGN performance on these test cases is further illustrated in Fig.~\ref{fig:Phy-MGN_vs_MGN}. For each model, we evaluate the mean squared error across the entire test dataset containing unseen 17 cases over all 100 predicted timesteps, with ambient density ranging from 3.2 to 19.2 gcc, providing a comprehensive assessment of temporal prediction accuracy over several unseen cases. The comparison shows that Phy-MGN yields a significantly better result than original MGN. With the addition of physics-informed loss, the model is biased towards conservation consistency. It appears to reduce autoregressive rollout drift, and improves long term stability.

\section{Phy-MGN performance on smooth and noisy datasets}
\label{app:D}

\begin{figure*} [htb!]
\vspace{-1em}
\includegraphics[width=0.9\textwidth]{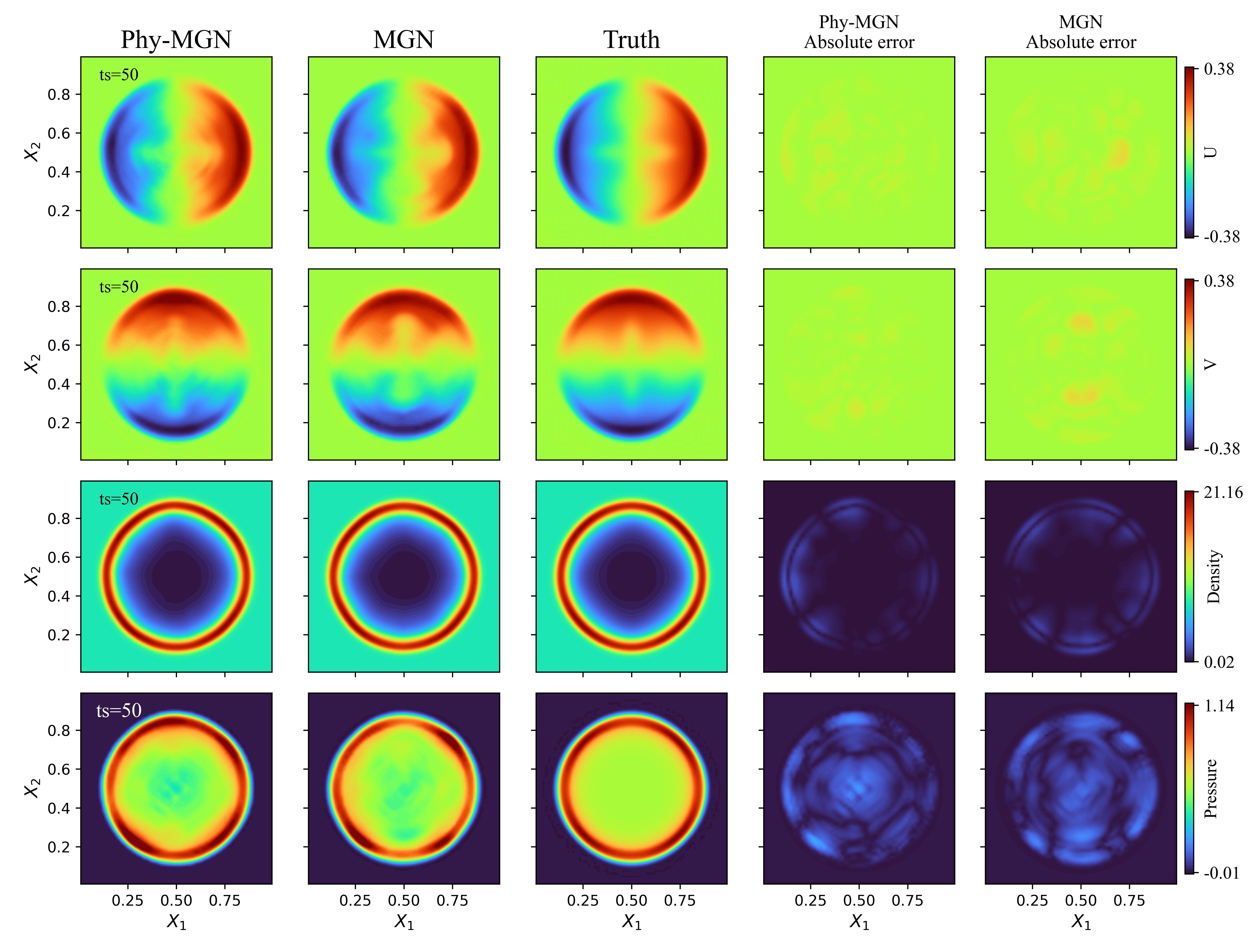}
\vspace{-1.5em} 
\includegraphics[width=0.9\textwidth]{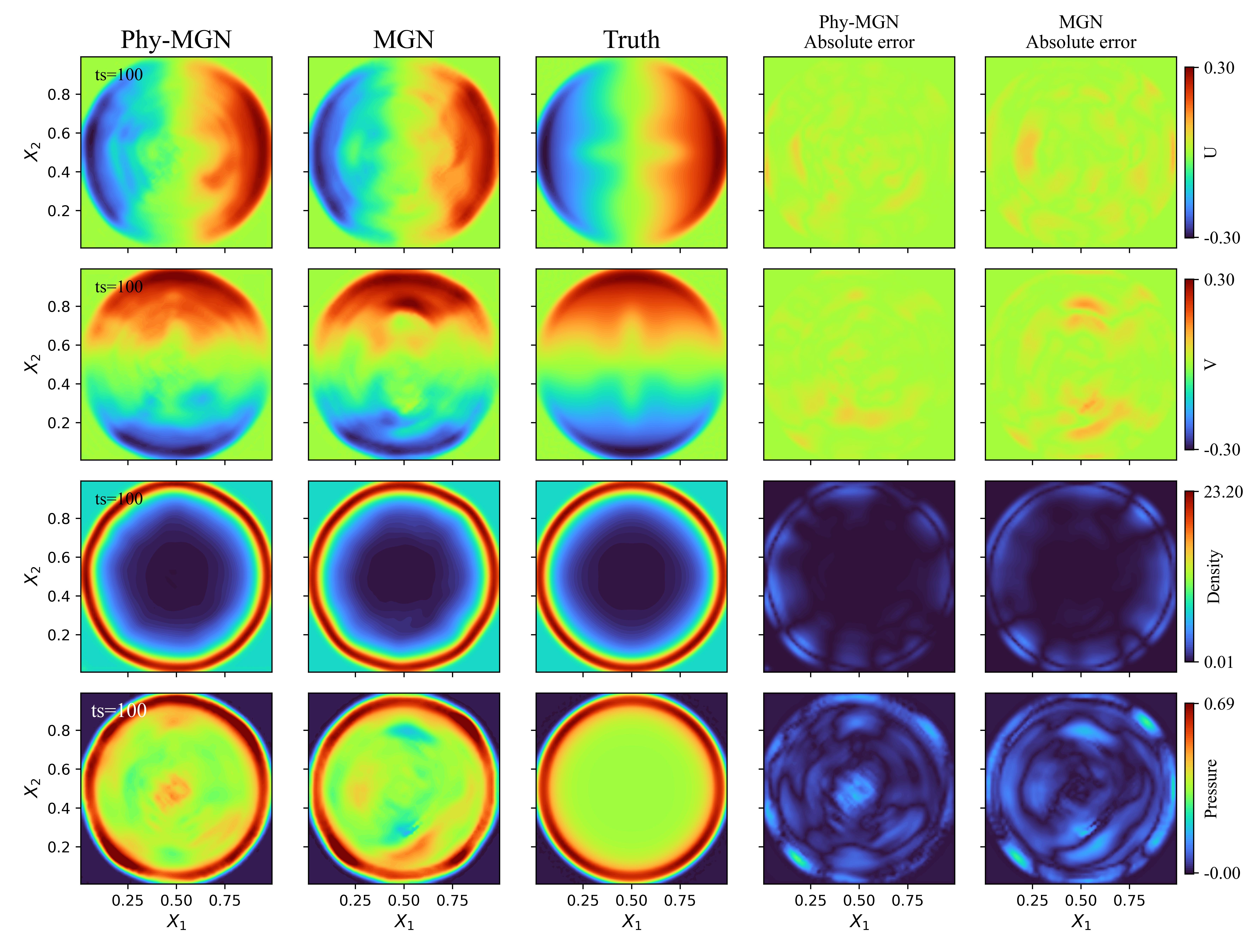}
\caption{\label{fig:rhoA_7.2} \textbf{Comparision of Phy-MGN and GNN prediction on a test case within training parameter space}. The figure above presents four physical states and corresponding losses for a case of ambient density 7.2 \(g/cm^3\), at the 50th (top panel) and 100th (bottom panel) inference timesteps. Model is trained using noisy dataset with numerical anisotropy, without smoothing algorithm.}
\end{figure*}

\begin{figure*} [htb!]
\vspace{-1em}
\includegraphics[width=0.9\textwidth]{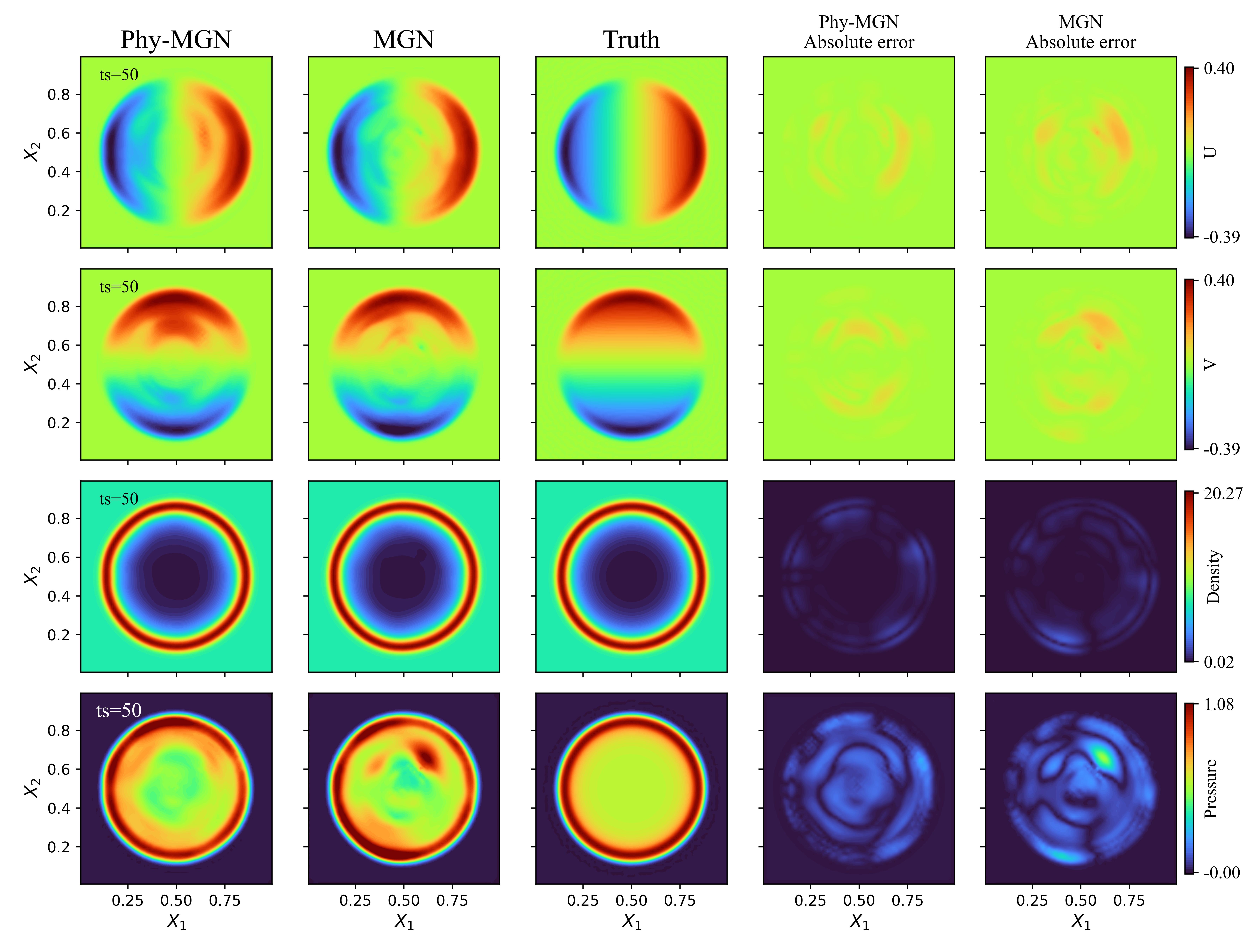}
\vspace{-1.5em} 
\includegraphics[width=0.9\textwidth]{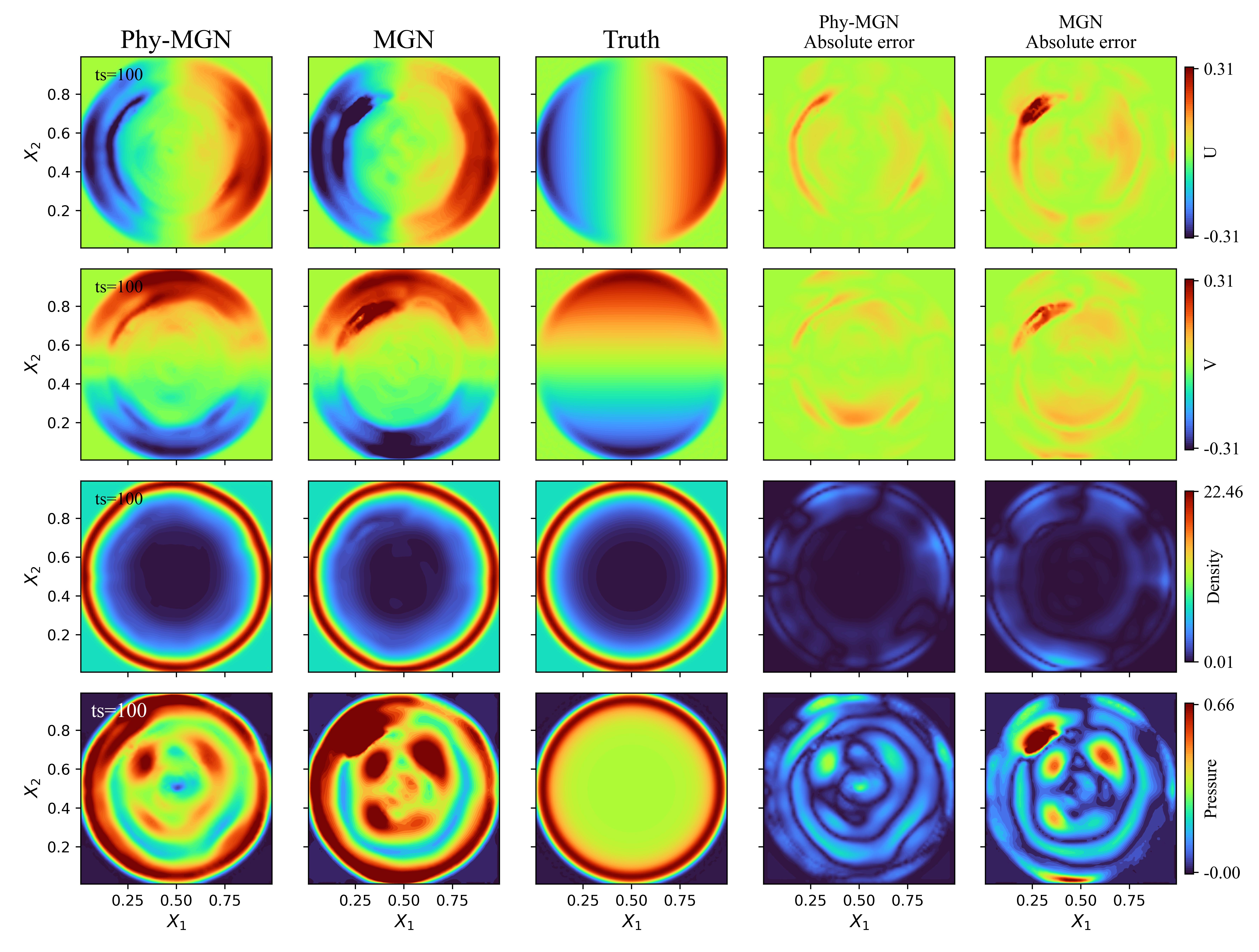}
\caption{\label{fig:rhoA_7.2_smooth} \textbf{Comparision of Phy-MGN and GNN prediction on a test case within training parameter space}. The figure above presents four physical states and corresponding losses for a case of ambient density 7.2 \(g/cm^3\), at the 50th (top panel) and 100th (bottom panel) inference timesteps. Model is trained using data processed by smoothing algorithm. }
\end{figure*}

\begin{figure*} [htb!]
\centering
\vspace{-1em}
\includegraphics[width=0.9\textwidth]{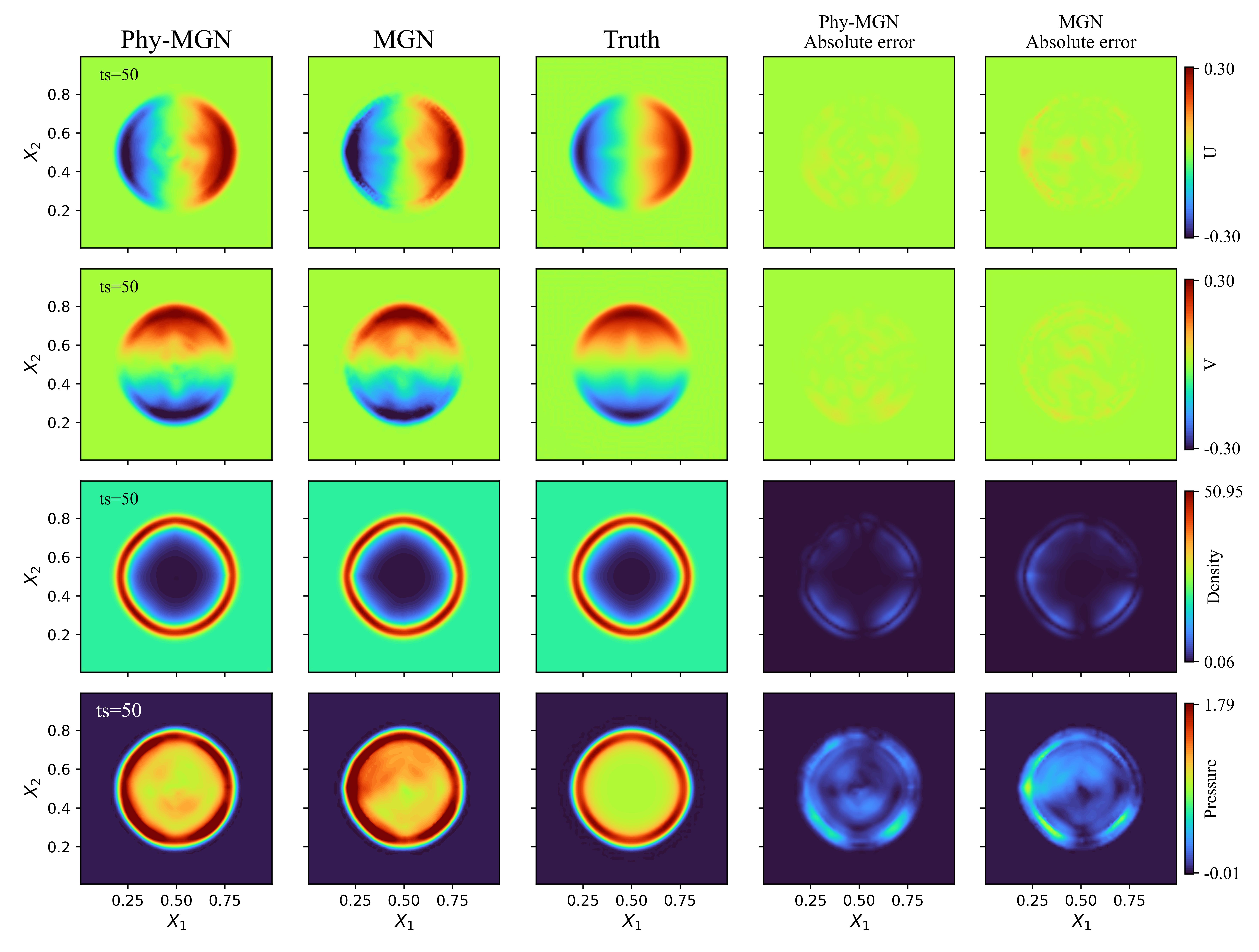}
\vspace{-1.5em} 
\includegraphics[width=0.9\textwidth]{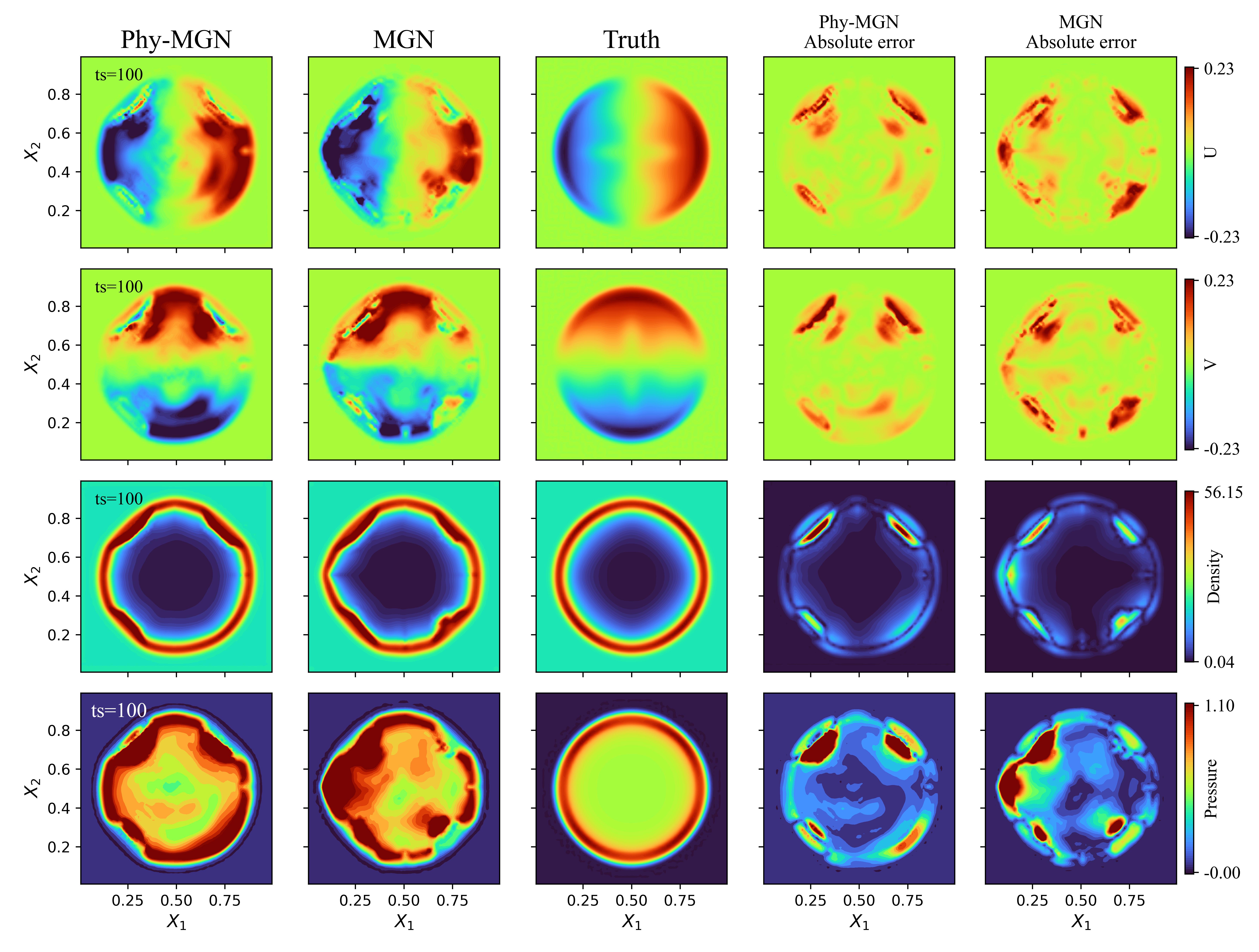}
\caption{\label{fig:rhoA_19} \textbf{Comparision of Phy-MGN and GNN prediction on a test case outside training parameter space}. The figure above presents four physical states and corresponding losses for a case of ambient density 19 \(g/cm^3\), at the 50th (top panel) and 100th (bottom panel) inference timesteps. Model is trained using noisy dataset with numerical anisotropy, without smoothing algorithm.}
\end{figure*}

\begin{figure*} [htb!]
\vspace{-1em}
\includegraphics[width=0.9\textwidth]{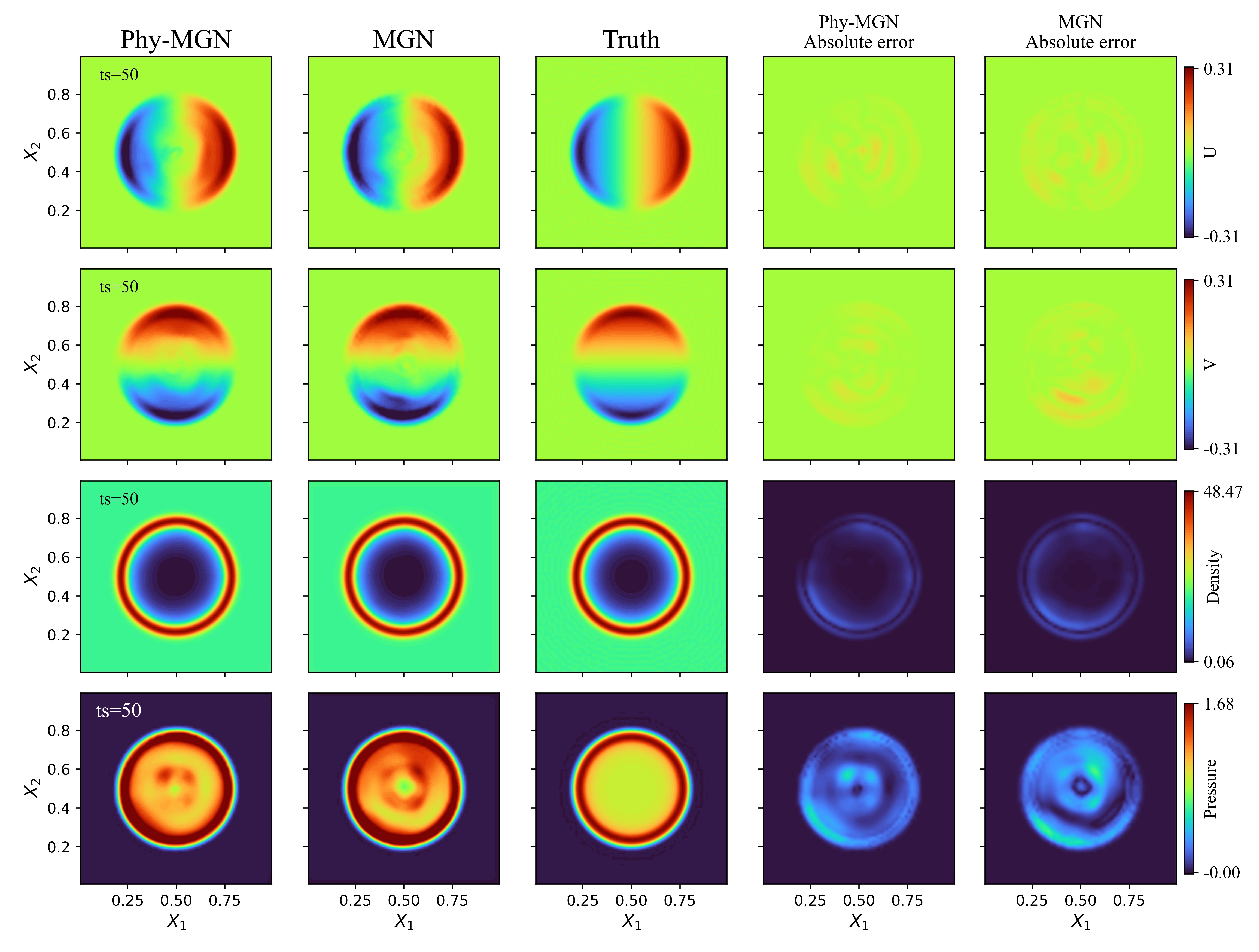}
\vspace{-1.5em} 
\includegraphics[width=0.9\textwidth]{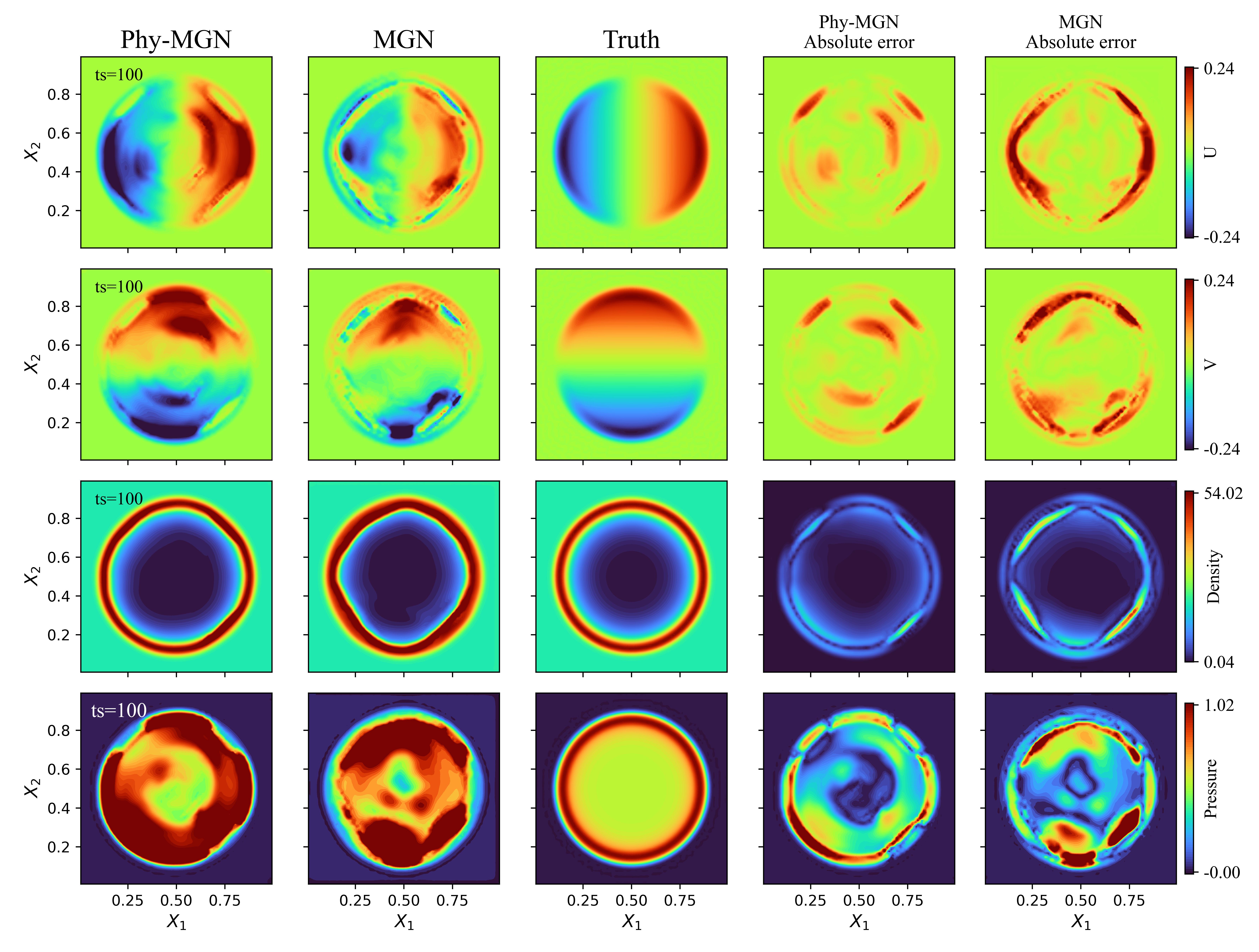}
\caption{\label{fig:rhoA_19_smooth} \textbf{Comparision of Phy-MGN and GNN prediction on a test case outside training parameter space}. The figures above present four physical states and corresponding losses for a case of ambient density 19 \(g/cm^3\), at the 50th (top panel) and 100th (bottom panel) inference timesteps. Model is trained using data processed by smoothing algorithm. }
\end{figure*}

\begin{figure*}[!htbp]
\includegraphics[width=\textwidth]{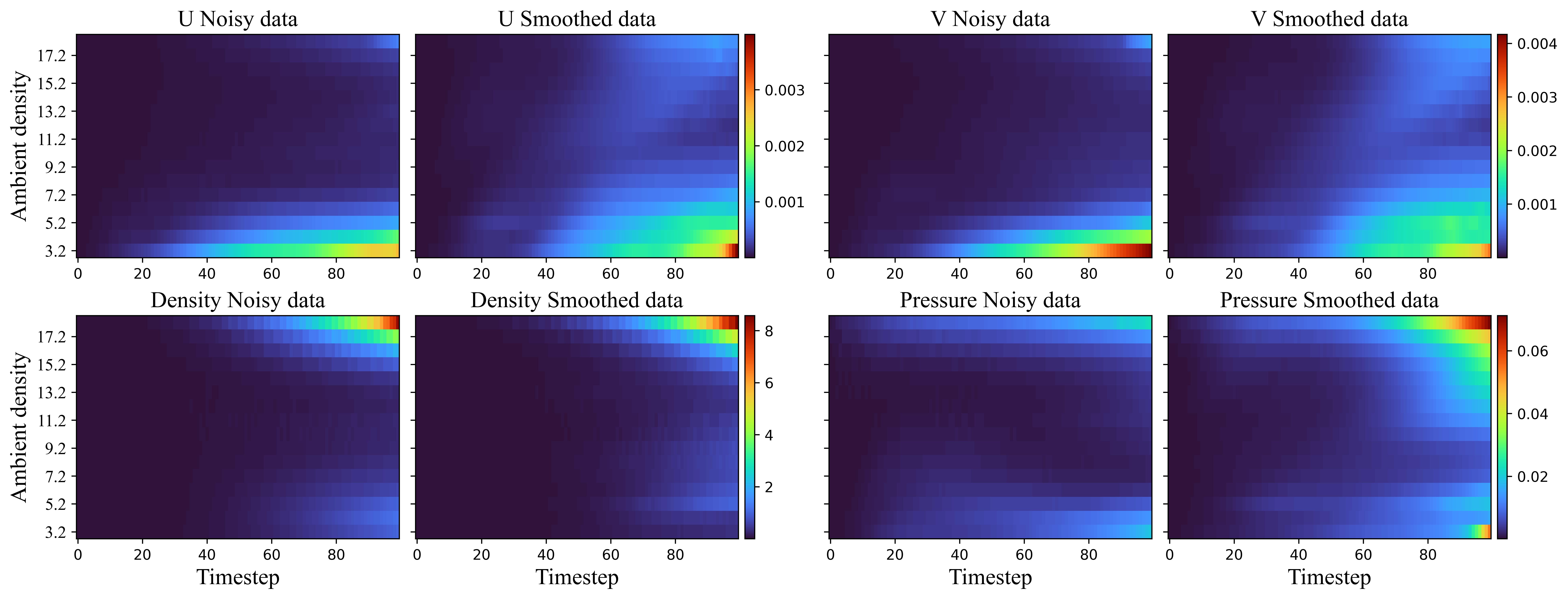}

\caption{\label{fig:noise_vs_smooth} \textbf{The mean squared error (MSE) of the Sedov problem} is plotted for results of Phy-MGN trained with noisy data (column 1 and 3) and smoothed data (column 2 and 4) respectively, across all four state variables: density, pressure, x-velocity U, and y-velocity V. Both model demonstrates a good prediction, with the model trained with noisy data having a better performance on pressure, which can be spotted by comparing visualizations above.}
\end{figure*}

In addition, we demonstrate that both Phy-MGN and MGN are capable of learning from datasets that contain nonphysical numerical noise. We compare training on two variants of the FLASH dataset—(i) raw simulation data that contain numerical artifacts generated by FLASH code and (ii) dataset where the numerical artifacts are reduced by a smoothing algorithm. In both cases, we additionally inject normalized Gaussian noise ($\sigma=0.2$) into the velocity fields to further stress the learning problem. The Phy-MGN results comparing (i) and (ii) for test case within (Fig.~\ref{fig:rhoA_7.2} and Fig.~\ref{fig:rhoA_7.2_smooth}) and outside (Fig.~\ref{fig:rhoA_19} and Fig.~\ref{fig:rhoA_19_smooth}) training parameter space show that Phy-MGN is not significantly affected by noise contamination, the robustness of the physics-informed approach. Figure.~\ref{fig:noise_vs_smooth} shows a more comprehensive assessment of the statement above, evaluating mean squared error across the entire test dataset containing unseen 16 cases over all 100 predicted timesteps. In this visual comparision, noise-contaminated dataset shows a slightly better performance than smoothed dataset on the prediction of pressure, while having no significant improvment on other quantities.


\begin{table*}[htbp]
\caption{\label{tab:quantitive} \textbf{Comparision of Phy-MGN and MGN result}. The Models are trained with smoothed data. The table displays the mean $\mu$ and the standard
deviation $\sigma$ of the mean square error calculated for 16 test cases for four physical states, in format $\mu \pm\sigma$. Phy-MGN always shows the best performance which is highlighted in blue.}
\begin{ruledtabular}
\begin{tabular}{cccccc}

 \emph{time step} & \emph{Model} & Density & Pressure & x-Velocity U & y-Velocity V   \\ \hline
 1 & Phy-MGN & \textcolor{blue}{\(1.3\!\times\!10^{-3} \pm 2.2\!\times\!10^{-3}\)}  & \textcolor{blue}{$2.1{\times}10^{-4} \pm 1.1{\times}10^{-4} $} & \textcolor{blue}{$5.7{\times}10^{-6} \pm 3.0{\times}10^{-6} $} & \textcolor{blue}{$6.0{\times}10^{-6} \pm 3.8{\times}10^{-6} $} \\ 
 1 & MGN & $1.3{\times}10^{-3} \pm 2.3{\times}10^{-3} $ & $2.5{\times}10^{-4} \pm 1.3{\times}10^{-4} $ & $1.1{\times}10^{-5} \pm 4.5{\times}10^{-6} $ & $1.2{\times}10^{-5} \pm 5.1{\times}10^{-6} $ \\ \hline
 50 & Phy-MGN & \textcolor{blue}{$1.1{\times}10^{-1} \pm 1.3{\times}10^{-1} $} & \textcolor{blue}{$3.6{\times}10^{-3} \pm 2.2{\times}10^{-3} $} & \textcolor{blue}{$2.9{\times}10^{-4} \pm 2.1{\times}10^{-4} $} & \textcolor{blue}{$2.9{\times}10^{-4} \pm 2.1{\times}10^{-4} $} \\
 50 & MGN & $1.9{\times}10^{-1} \pm 1.5{\times}10^{-1} $ & $8.8{\times}10^{-3} \pm 3.9{\times}10^{-3} $ & $4.9{\times}10^{-4} \pm 3.0{\times}10^{-4} $ & $4.8{\times}10^{-4} \pm 2.8{\times}10^{-4} $\\ \hline
 100 & Phy-MGN & \textcolor{blue}{$1.5\pm 2.1 $} &\textcolor{blue}{ $2.5{\times}10^{-2} \pm 1.8{\times}10^{-2} $} & \textcolor{blue}{$8.8{\times}10^{-4} \pm 9.7{\times}10^{-4} $} & \textcolor{blue}{$8.7{\times}10^{-4} \pm 7.1{\times}10^{-4} $} \\
 100 & MGN & $3.5\pm 3.9 $ & $7.4{\times}10^{-2} \pm 6.4{\times}10^{-2} $ & $2.1{\times}10^{-3} \pm 2.2{\times}10^{-3} $ & $1.9{\times}10^{-3} \pm 1.7{\times}10^{-3} $  \\


\end{tabular}
\end{ruledtabular}
\end{table*}

\section{Spatial-Temporal error distribution of training data}
\label{app:E}
The optimization process in model training aims to minimize the MSE loss that combines both the data driven loss and the PDE loss. In practice, however, these two loss terms do not always decrease simultaneously. This mismatch arises from several sources of irreducible error inherent to simulation-generated datasets : (i) numerical instabilities and scheme-dependent artifacts in the simulation code introduce nonphysical noise into the data (ii) the imperfect numerical discretization used to evaluate the PDE residual also introduces irreducible error, since the governing equations are enforced through finite-difference or finite-volume approximations rather than in their continuous form. (iii) and the unknown physics  that we did not include in the conservation equations. Together, these factors prevent the PDE residual from vanishing even for the ground-truth solution. For this reason, the physics-informed loss cannot be formulated using the absolute PDE residual alone; instead, it must be defined relative to the residual present in the ground-truth data so that both the data loss and physics-informed loss can be minimized consistently during training.

\begin{figure*}[htbp]
\includegraphics[width=0.8\textwidth]{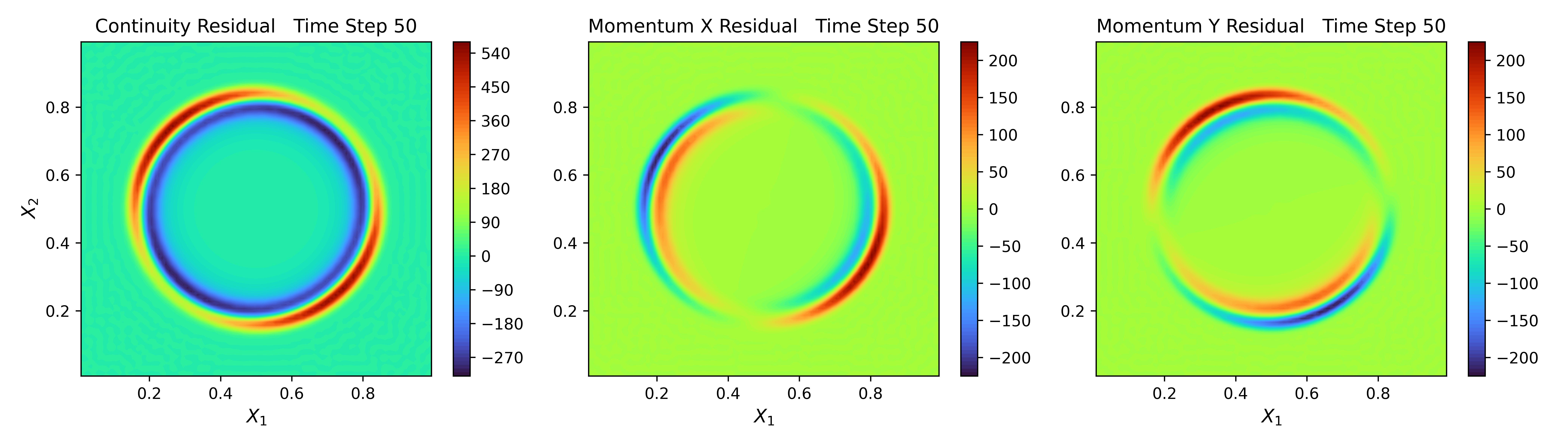}
\caption{\label{fig:Sedov res} \textbf{Spatial–temporal distributions of the residuals of the mass and momentum conservation equations for a representative Sedov–Taylor explosion test case}. The residuals are evaluated from the simulation output using finite-difference scheme. The regions of largest deviation occur near the expanding shock front, indicating where the numerical solution most strongly departs from ideal conservation and highlighting where unresolved physics and discretization errors dominate. The high residual value is caused by the small spatial and temporal step sizes in the denominator.}
\end{figure*}

Here we show a visualisation of the spatialtemporal distribution of the PDE residual present in ground truth in Fig.~\ref{fig:Sedov res}. But showing where the error is the most significant. The visualisation shows that the shock front are usually the area with the highest PDE residual in ground truth data. As in these highly compressed region has sharp discontinuities and the continueous Euler equation is no longer valid in these region. For some cases, the value of PDE residuals are extremely high --- this is caused by dividing small dt and dx when calculating PDE residuals,  small dt and dx will further amplify these errors, to address this, we multiply the PDE residual by dt, for them to be on the same unit as data loss. 
The results show that the largest PDE residuals are typically concentrated near shock fronts. In these highly compressed regions, sharp discontinuities arise, and the assumptions underlying the continuous Euler equations are no longer strictly valid, leading to elevated residuals even in the ground-truth solution. In some cases, the PDE residual values become extremely large due to division by small temporal and spatial step sizes \(\Delta t\) and \(\Delta x\) during residual evaluation, which further amplifies local numerical errors. To mitigate this effect and maintain consistency with the data-driven loss, we scale the PDE residual by multiplying with \(\Delta t\), ensuring that the physics-informed loss is expressed in the same unit to the data loss.

\clearpage
\bibliography{ref}
\end{document}